\documentclass{article}
\usepackage{graphicx}  
\usepackage{amsmath}   
\usepackage[compress,numbers,sort]{natbib}
\usepackage{amssymb}   
\usepackage{bm} 
\usepackage{dcolumn}
\usepackage{color}
\usepackage{mathrsfs}
\usepackage{amsfonts}
\usepackage{varioref}
\usepackage{textcomp}
\usepackage[normalem]{ulem}

\RequirePackage[colorlinks,citecolor=blue,urlcolor=magenta,linkcolor=blue]{hyperref}
\allowdisplaybreaks
\labelformat{section}{Section #1} 
\labelformat{subsection}{Section #1} 
\labelformat{subsubsection}{Section #1}
\labelformat{subsubsubsection}{Section #1}
\labelformat{equation}{Eq.~(#1)} 
\labelformat{figure}{Fig.~#1} 
\labelformat{subfigure}{Fig.~\thefigure#1} 
\labelformat{table}{Table~#1} 
\labelformat{appendix}{Appendix #1}
\title{\bf Addressing issues in defining the Love numbers for black holes}
\author{Rajendra Prasad Bhatt\footnote{rajendra@iucaa.in}$~^{1}$\href{https://orcid.org/0009-0004-9088-2998}{\includegraphics[scale=0.07]{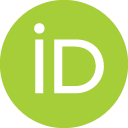}}, Sumanta Chakraborty\footnote{tpsc@iacs.res.in}$~^{2}$\href{https://orcid.org/0000-0003-3343-3227}{\includegraphics[scale=0.07]{ORCIDiD_icon128x128.png}}, and Sukanta Bose\footnote{sukanta@wsu.edu}$~^{1,3}$\href{https://orcid.org/0000-0002-4151-1347}{\includegraphics[scale=0.07]{ORCIDiD_icon128x128.png}}
\\
$^{1}${\small{Inter-University Centre for Astronomy and Astrophysics,}}\\
{\small{Pune 411007, India}}\\
$^{2}${\small{School of Physical Sciences,}}\\
{\small{Indian Association for the Cultivation of Science, Kolkata-700032, India}}\\
$^{3}${\small{Department of Physics and Astronomy,}}\\
{\small{ Washington State University, 1245 Webster, Pullman, Washington 99164-2814, USA}}\\
}
\begin{document}
  
\maketitle
\begin{abstract} 
We present an analytic method for calculating the tidal response function of a nonrotating and a slowly rotating black hole from the Teukolsky equation in the small-frequency and near horizon limit. We point out that in the relativistic context, there can be two possible definitions of the tidal Love numbers and the dissipative part that arises from the tidal response function. Our results suggest that both of these definitions predict zero tidal Love numbers for a nonrotating black hole. On the other hand, for a slowly rotating black hole in a generic tidal environment, these two definitions of the tidal Love numbers do not coincide. While one procedure suggests zero tidal Love numbers, the other procedure gives purely imaginary tidal Love numbers. As expected, the dissipative terms differ as well. We emphasize that in our analysis, we keep all the terms linear in the frequency, unlike previous works in the literature. Following this, we  propose a procedure to calculate the tidal response function---and hence the Love numbers---for an arbitrarily rotating black hole. 

\end{abstract}

\section{Introduction}
Einstein’s general theory of relativity makes 
exciting predictions 
about cosmology and various
astrophysical objects---with
black holes, 
arguably, as
the simplest and most important among them. 
Recent observations
of gravitational waves by the LIGO-Virgo Collaboration~\cite{LIGOScientific:2016aoc, LIGOScientific:2021djp}, and captures 
of
the images of shadows cast by black holes  
with 
the Event Horizon Telescope~\cite{EventHorizonTelescope:2019dse, EventHorizonTelescope:2022wkp} are 
strongly
suggestive of their existence.
Among 
the
very many properties of
these compact objects
that
have been studied in the past, the response of a black hole to a tidal environment is 
an especially
exciting
one
since it can 
form the basis of
a smoking-gun test of the black hole nature of 
compact objects. In this article, we study this property and
provide an analytic method for computing the tidal response of a black hole to an external tidal field in a relativistic scenario.

The response of any self-gravitating object under the influence of an external tidal field can be divided into two parts---conservative and dissipative. The conservative part encodes the information about the tidal deformation of the body, while the dissipative part describes the absorption of the emitted gravitational waves by the deformed object \cite{Chia:2020yla}. One can associate dimensionless numbers with the body's tidal deformation~\cite{Hinderer:2007mb, Flanagan:2007ix, Damour:2009vw, Binnington:2009bb}, known as the tidal Love numbers. Several studies, including Newtonian as well as relativistic, have found the tidal Love numbers for a Schwarzschild black hole to be zero~\cite{Binnington:2009bb, Damour:2009vw, Kol:2011vg, Chakrabarti:2013lua, Gurlebeck:2015xpa, LeTiec:2020spy, LeTiec:2020bos, Chia:2020yla, Charalambous:2021mea, Hui:2020xxx, Creci:2021rkz}. Generalization of this result 
to include slowly rotating black holes has also been attempted before, in the context of Newtonian gravity~\cite{Landry:2015zfa, Pani:2015hfa}, and zero tidal Love numbers 
have been obtained there as well. The corresponding situation for an arbitrarily rotating black hole has remained ambiguous. Attempts have also been made to study the nonlinearities~\cite{DeLuca:2023mio} and stability~\cite{Katagiri:2023yzm} of the tidal response of a Schwarzschild black hole. The analysis of the tidal Love numbers is generically performed using the Newtonian approximation for the perturbed $g_{tt}$ component of the metric and then expanding the perturbation asymptotically. However, this method lacks the covariant nature of general relativity, such that a different choice for the asymptotic coordinates can alter the numerical value of the tidal Love numbers. Thus, a relativistic generalization has been much sought after. 

To further stir the already muddled situation, \cite{LeTiec:2020spy, LeTiec:2020bos} reported nonzero but imaginary tidal Love numbers for a rotating black hole in a nonaxisymmetric tidal environment, in complete contrast with all the previous works in the literature. At the same time, \cite{LeTiec:2020spy, LeTiec:2020bos} also presented a covariant method of determining the tidal Love numbers namely through the Weyl scalar---and the tidal Love numbers in the zero-frequency limit were obtained by first solving the radial Teukolsky equation and then taking its asymptotic limit. Though the result was counterintuitive, \cite{LeTiec:2020spy, LeTiec:2020bos} set the stage for determining the relativistically invariant tidal Love numbers. Based on this suggestion, \cite{Chia:2020yla} computed the tidal response function from the radial Teukolsky equation under the small-frequency approximation from the asymptotic expansion of the Weyl scalar and demonstrated that the imaginary Love numbers calculated in \cite{LeTiec:2020spy, LeTiec:2020bos} are actually associated with the dissipative part of the response function, and not with the conservative part. Thus, it turned out that the tidal Love numbers of both nonrotating and rotating black holes identically vanish. In this work, we revisit the analysis and discuss possible subtleties associated with the definition of the tidal Love numbers, which earlier works have missed. In particular, we will focus on extracting the tidal Love numbers from the response function of black holes under an external tidal field and shall also closely study the approximations involved in solving the radial Teukolsky equation.  

At this outset, it is also worthwhile to point out the implications of the analysis involving tidal Love numbers. If the black holes in general relativity indeed have zero tidal Love numbers, then not only can we test the black hole paradigm, but we will also be able to distinguish black holes in general relativity from those in alternative theories of gravity. This is because compact objects other than black holes e.g., wormholes, boson stars, and gravastars---all have non-zero tidal Love numbers~\cite{Yagi:2016ejg,Yazadjiev:2018xxk,Cardoso:2017cfl,Chakravarti:2019aup}; even black holes with quantum corrections have nonzero Love numbers~\cite{Nair:2022xfm}. Thus, from the inspiral part of the gravitational wave signal arising from the coalescence of two compact objects, if one can read off the contribution from the tidal Love numbers, then it will be possible to comment on the nature of these compact objects. Moreover, the same test can also be used to distinguish general relativity from other theories of gravity, as black holes in 
some alternative theories of gravity
may have nonzero Love numbers~\cite{DeLuca:2022tkm}, in contrast to Schwarzschild or Kerr black holes. For example, higher-dimensional black holes have nonzero tidal Love numbers~\cite{Chakravarti:2018vlt}. Thus, with the advent of the new generation of gravitational wave detectors, the question regarding the nonzero value of the tidal Love numbers for compact objects can possibly be answered and will be of much significance. As emphasized already, an understanding of the tidal Love numbers is useful not only for a more complete comprehension of black hole physics at the classical level, but also for probing a wider variety of compact objects and the fundamental physics associated with them. For instance, nonzero tidal Love numbers may shed light on the quantum nature of gravity, besides providing a better understanding of the existence of exotic compact objects and various properties of their constituents.

The present work is organized as follows: We start in \ref{response_tidal} by reviewing the definition of the tidal response function in the Newtonian context and its generalization to the relativistic domain. Subsequently, we  provide possible definitions of the tidal Love numbers arising from the response function and elaborate on the motivations behind them. In \ref{small_freq_master}, we  discuss the approximations involved and hence present the master equation, obtained from the radial Teukolsky equation, which will be used to calculate the response function. The computation of the response function of the Schwarzschild black hole in an external tidal field has been presented in \ref{response_Schwarzschild} and the corresponding expression for the slowly rotating Kerr black hole is depicted in \ref{response_slow}. Finally, in \ref{discussion} we discuss possible methods of determining the tidal response function for an arbitrarily rotating black hole in light of the results obtained in this work and also suggest possible future directions of exploration. We have delegated several detailed computations in the appendixes. 

\emph{Notations and conventions:} In this work, we have set the relevant fundamental constants to unity, i.e.,  $G=1=c$, unless otherwise stated. We will mostly use the positive signature convention, such that the flat spacetime Minkowski metric in the Cartesian coordinates takes the following form: $\eta_{\mu \nu}=\textrm{diag.}(-1,+1,+1,+1)$. Moreover, the greek letters $\mu,\nu,\alpha,\ldots$ denote the four spacetime coordinates, while the lowercase roman letters $i,j,k,\ldots$ denote the three-dimensional space coordinates. The boldfaced roman letters, e.g., ${\bm A}$, will denote three-dimensional (spatial) vectors.

\section{Tidal response function in a relativistic setting}\label{response_tidal}

In this section, we will first briefly review the response of a compact object to the tidal field of a companion in the Newtonian context. We will subsequently describe how the response function can be derived in a relativistic setting through the Newman-Penrose formalism~\cite{Newman:1961qr}. The generalization of the response function to the relativistic setting is necessary since the response function in the Newtonian regime is prone to issues related to the choice of gauge.
Afterwards, we will demonstrate how an expression for the Love numbers can be arrived at from the relativistic formulation of the tidal response function.
\subsection{Newtonian response function}
Consider a spherically symmetric, nonrotating (or slowly rotating) compact object of mass $M$ immersed in an external gravitational field, which acts as the source of tidal perturbation. The total gravitational potential outside the object will be the combination of the gravitational potential due to the object itself, given by $U_{\text{body}}$, and the potential of the tidal background, denoted by $U_{\text{tidal}}$. 
Among these, the gravitational potential $U_{\text{body}}$ can be expressed in terms of the mass multipole moments of the gravitating object and is given by a series expansion in inverse powers of the radial coordinate $r$ as~\cite{poisson_will_2014, LeTiec:2020bos}, 
\begin{equation}
U_{\text{body}}(t,{\bm x}) = \sum_{l=0}^{\infty}\frac{(2l-1)!!}{l!}I^{\langle L\rangle}\frac{n_{\langle L\rangle}}{r^{l+1}}~.
\label{Uext}
\end{equation}
In the above expression, $L$ as a superscript denotes the multi-index set $a_1 a_2\cdots a_l$, such that $I^L \equiv I^{a_1 a_2\cdots a_l}$, describes a tensor of rank $l$, and $I^{\langle L\rangle}$ denotes the symmetric trace-free (STF) part of it~\cite{poisson_will_2014}. For illustration, we present below the symmetric and trace-free part of a second-rank tensor $A^{jk}$:
\begin{equation}
A^{\langle jk\rangle} \equiv \frac{1}{2}\left(A^{jk} + A^{kj}\right) - \frac{1}{3}\delta^{jk}A^i{}_i~,
\end{equation}
and similar definitions exist for the STF part of any higher-rank tensors as well. Along similar lines, the tensor $n^{L} \equiv x^L/r^l$, where $x^L = x^{a_1 a_2\cdots a_l} \equiv x^{a_1}x^{a_2}\cdots x^{a_l}$ and $r=\sqrt{x^{i}x_{i}}$, with $n^{\langle L\rangle}$ being the STF part of $n^L$. Note that $n^{L}$ is simply a direct product of $l$ unit vectors, each of which is defined as ${\bm n}\equiv {\bm x}/r$, where $\bm{x}$ is the spatial vector with respect to the Galilean transformation in the Cartesian coordinates. The symmetric and trace-free tensor $n^{\langle L\rangle}$ can also be expressed in terms of the spherical harmonics as follows~\cite{poisson_will_2014}:
\begin{equation} 
n^{\langle L\rangle}:=\frac{4\pi l!}{(2l+1)!!}\sum_{m=-l}^{l}\mathscr{Y}_{lm}^{\langle L\rangle}Y_{lm}(\theta, \phi)~,
\qquad
Y_{lm}(\theta,\phi)=\mathscr{Y}_{lm}^{*\langle L\rangle}n_{\langle L\rangle}~,
\end{equation}
where $\mathscr{Y}_{lm}^{\langle L\rangle}$ is a constant STF tensor, relating the components of the normal STF tensor with the spherical harmonics. 

The gravitational potential of the compact object also depends on the STF tensor $I^{\langle L\rangle}$, which is related to its mass multipole moments. In particular, the $l=0$ term in \ref{Uext} falls off as $r^{-1}$, and hence corresponds to the monopole mass moment, which depends on the mass of the compact object. In contrast, the higher values of $l$ in the expansion of $U_{\rm body}$ in negative powers of the radial coordinate are related to the higher mass multipole moments of the compact object.\footnote{The $l=1$ term will be zero, when we place the origin at the center of mass of the object.} These mass multipole moments, namely $I^{\langle L\rangle}$, are defined in terms of an integral over the volume of the compact object as
\begin{equation}
I^{\langle L\rangle}= \int \rho(t,{\bm x}) x^{\langle L\rangle} \mathrm{d}^3{\bm x}~,
\end{equation}
where the STF tensor $x^{\langle L\rangle}$ has already been introduced earlier, and $\rho(t,{\bm x})$ is the mass density of the compact object, which also includes any perturbations to the mass density created by the tidal field. Due to the time variation of the mass density, these mass multipole moments, described by the STF tensors $I^{\langle L\rangle}$, are also functions of the time coordinate. For our purpose, it will be convenient to transform these STF tensors to the spherical harmonic basis---e.g., $I^{\langle L\rangle}$ should be transformed to $I_{lm}$, which are the spherical harmonic modes of the multipole moments. Such a transformation reads~\cite{LeTiec:2020bos,poisson_will_2014}
\begin{equation}
I^{\langle L\rangle} = \sum_{m=-l}^{l}\mathscr{Y}_{lm}^{*\langle L\rangle}I_{lm}~,
\label{T1}
\end{equation}
while the inverse transformation, expressing $I_{lm}$ in terms of the STF multipole moment tensor $I^{\langle L\rangle}$, yields
\begin{equation}
I_{lm} = \frac{4\pi l!}{(2l+1)!!}\mathscr{Y}_{lm}^{\langle L\rangle}I_{\langle L\rangle}~.
\label{T2}
\end{equation}
Note that the normalization factors in both of these relations are arbitrary, and different conventions exist for them. For example, the choice of the normalization factors adopted here corresponds to \cite{LeTiec:2020bos}, while in \cite{poisson_will_2014} the factor of $\{4\pi l!/(2l+1)!!\}$ appearing on the right-hand side of \ref{T2} is interchanged with the unit factor of \ref{T1}. 

In an analogous manner, one can also express the gravitational potential due to the tidal field as a series in the positive powers of the radial coordinate $r$ whose coefficients denote the tidal moments, such that
\begin{equation}
U_{\text{tidal}}(t,{\bm x}) = -\sum_{l=2}^{\infty}\frac{(l-2)!}{l!}\mathcal{E}_{\langle L\rangle}(t)n^{\langle L\rangle}r^{l}~,
\label{Utidal}
\end{equation}
where one defines the time-dependent tidal moments as
\begin{equation}
\mathcal{E}_{\langle L\rangle}(t)=-\frac{1}{(l-2)!}\partial_{\langle L\rangle}U_{\text{tidal}}\Big\vert_{t,{\bm x}=\bm{0}}~.
\label{el}
\end{equation}
It should be emphasized that \ref{Utidal} should be considered as a Taylor expansion in the ratio $(r/\mathcal{L})$, where $r$ is the distance of the field point from the center of mass of the deformed object and $\mathcal{L}$ is a characteristic length scale over which the tidal field varies in space. In particular, $\mathcal{E}_{\langle L\rangle}\sim \mathcal{L}^{-l}$, and we will assume that the tidal field varies slowly over space, so that over the region of interest, the condition $(r/\mathcal{L})\ll1$ is always satisfied~\cite{LeTiec:2020bos}.

Besides, in the above expression, $\partial_{L}U_{\text{tidal}}$ corresponds to the $l$th derivative of $U_{\text{tidal}}$ evaluated at the center of mass of the body, located at $\bm{x}=\bm{0}$, and $\partial_{\langle L\rangle}U_{\text{tidal}}$ corresponds to the associated STF tensor. Like the multipole moments, the spherical harmonic components of the tidal field and the quantities $\mathcal{E}_{\langle L\rangle}(t)$ are also related through \ref{T1}, while the inverse relation is obtained with $I$ replaced by $\mathcal{E}$ in \ref{T2}. We will have occasion to use these relations subsequently. Again, the overall normalization factor in \ref{Utidal}, as well as in \ref{el}, can differ depending on the convention employed. For example, the choice considered here is consistent with \cite{LeTiec:2020bos, Chia:2020yla}. On the other hand, in \cite{poisson_will_2014} the term involving $(l-2)!$ is absent in the normalization of both \ref{Utidal} and \ref{el}. However when the tidal moments $\mathcal{E}_{\langle L\rangle}$ from \ref{el} are substituted back into the tidal potential in \ref{Utidal}, it is evident that the $(l-2)!$ term appearing in the present situation gets canceled and the resulting equation will coincide with the corresponding expression in \cite{poisson_will_2014}. 

Having elaborated on the spherical harmonic decomposition of the Newtonian potential associated with the compact object and also with the tidal field, we find for the total Newtonian potential the following decomposition:
\begin{equation}
U=U_{\text{body}}+U_{\text{tidal}} = \frac{M}{r} + \sum_{l=2}^{\infty}\sum_{m=-l}^{l}\left[\frac{(2l-1)!!}{l!}\frac{I_{lm}Y_{lm}}{r^{l+1}}-\frac{(l-2)!}{l!}\mathcal{E}_{lm}Y_{lm}r^l\right]~,
\label{potential}
\end{equation}
where we have separated out the monopole term. Since the higher-order mass multipole moments, $I_{lm}$, are generated due to the external tidal field $\mathcal{E}_{lm}$, in the linear regime these should be proportional to each other. This suggests the following linear response of the gravitating compact object to a slowly varying tidal field \cite{Chia:2020yla}:
\begin{equation}
I_{lm} = -\frac{(l-2)!}{(2l-1)!!} R^{2l+1} \left[2k_{lm} \mathcal{E}_{lm} -\tau_0 \nu_{lm} \dot{\mathcal{E}}_{lm} + \cdots\right]~,
\label{def_I}
\end{equation}
where $-\{(l-2)!/(2l-1)!!\}$ is the overall normalization factor\footnote{Readers may find other choices for this normalization factor in the literature. It depends on the definitions of the potentials, the mass multipole moments, and their transformation to the corresponding STF counterparts.} and $R$ is a characteristic length scale, often taken to be the radius of the compact object that is being tidally deformed. The quantity $k_{lm}$, appearing as the proportionality factor between the mass multipole moments and the tidal field, is referred to as the tidal Love numbers, and its determination will be one of our primary focuses in this paper. On the other hand, the connection between the multipole moment and the time derivative of the tidal field is defined by the term $\tau_0\nu_{lm}$, where $\tau_{0}$ is a characteristic timescale over which the tidal field changes significantly over time and $\nu_{lm}$ is the proportionality factor that we relate to the tidal dissipation. Substituting \ref{def_I} into \ref{potential}, and moving to the Fourier space, we obtain the following expression for the total gravitational potential:
\begin{equation}
U = \frac{M}{r} - \sum_{l=2}^{\infty}\sum_{m=-l}^{l}\frac{(l-2)!}{l!}\mathcal{E}_{lm}r^l\left[1+F_{lm}(\omega)\left(\frac{R}{r}\right)^{2l+1}\right]Y_{lm}~,
\label{potential_Fourier_space}
\end{equation}
where $\omega$ is the mode frequency and we have introduced the  function 
\begin{equation}
F_{lm}(\omega)\equiv 2k_{lm}  +i\omega\tau_0 \nu_{lm} + \mathcal{O}(\omega^{2})~,
\label{def_resp}
\end{equation}
defined as the tidal response function in the Fourier space~\cite{Chia:2020yla}. 
Therefore, the real part of the response function, $k_{lm}$, is the tidal Love numbers, and the imaginary part, $\tau_0\nu_{lm}$, which is also proportional to the frequency $\omega$, is associated with the tidal dissipation. This formalism can also be generalized to a slowly rotating object~\cite{LeTiec:2020bos, Charalambous:2021mea}, where the structure of the gravitational potential remains the same, with the time derivative of the tidal field being with respect to the time coordinate in the corotating frame of reference. Therefore, the tidal response function becomes
\begin{equation}
F_{lm}(\omega')=2k_{lm}+i\omega'\tau_0 \nu_{lm} + \mathcal{O}(\omega'^{2})~,
\label{def_resp_sr}
\end{equation}
where $\omega'$ is the mode frequency in the body's corotating frame~\cite{poisson_will_2014}. Here too, the real part corresponds to the tidal Love numbers, and the imaginary part, proportional to the corotating frequency, is related to tidal dissipation.

However, this approach to determining the tidal response function depends heavily on the asymptotic expansion of the gravitational potential in the presence of an external tidal field. In the metric formulation, one relates the Newtonian potential to the time-time component of the metric, such that $g_{tt}=-(1-2U)$. Thus, the asymptotic expansion of the metric perturbation of the time-time component of the metric is related to the higher multipole moment expansion of the Newtonian potential, as in \ref{potential}. Therefore, using the technique outlined above, one can calculate the tidal response function from the time-time component of the metric perturbation. In general, these computations are Newtonian in nature and are dependent on the gauge choices---e.g., in the nonrotating case the tidal response is determined in the Zerilli gauge. As a consequence, this approach to calculating the tidal response function depends explicitly on the choice of coordinates and/or gauges and hence makes the interpretation of the results involving tidal Love numbers ambiguous~\cite{Gralla:2017djj}. Following this, we wish to present a covariant approach to determining the response function, and hence the Love numbers. 
\subsection{Tidal response function using the Newman-Penrose formalism}

In the previous section, the Newtonian definition of the tidal Love numbers through multipole expansion and spherical harmonic decomposition of the potentials was presented. However, as emphasized above, the Newtonian definitions have inherent gauge/coordinate ambiguities that can affect the numerical estimate of the tidal Love numbers values and are undesirable. In this work, we are interested in a relativistically invariant definition of the tidal Love numbers and, at the same time, we seek its imprint in gravitational waves arising from the object and propagating outward to future null infinity. These requirements single out the Weyl scalar $\psi_4$ as the starting point of deriving the relativistic tidal response. However, as a first step toward establishing $\psi_{4}$ as the main character behind the determination of the relativistic tidal Love numbers, we must demonstrate the connection between the Newman-Penrose scalar $\psi_4$ and the Newtonian potential, so that the corresponding result in the Newtonian limit can be derived. For this purpose, we can compute the Weyl scalar $\psi_{4}$ for the Newtonian metric, whose components in the Cartesian coordinate system read~\cite{LeTiec:2020bos}
\begin{equation}\label{Newt_metric}
g_{00} = -1 + \frac{2U}{c^2} + \mathcal{O}(c^{-4})~;
\qquad 
g_{0i} = \mathcal{O}(c^{-3})~;
\qquad 
g_{ij} = \delta_{ij}\left(1 + \frac{2U}{c^2}\right) + \mathcal{O}(c^{-4})~,
\end{equation}
where $U$ is the total Newtonian potential of the body and of the tidal field, together. Given the above metric, we can calculate the relevant components of the Weyl tensor using the Kinnersley tetrad \cite{Kinnersley:1969zza}, which in the Cartesian coordinate system read
\begin{equation}
l^\alpha =(1,n^i)~;\quad n^\alpha =\frac{1}{2}(1,-n^i)~;\quad m^\alpha = (0,m^i)~;\quad \Bar{m}^\alpha = (0,\bar{m}^i)~,
\end{equation}
where ${\bm n}={\bm x}/r$ is the normal three-vector defined earlier and $\bar{\bm m}$ is best expressed in the spherical polar coordinate system: $\bar{\bm m} =(1/\sqrt{2}r)(\partial_\theta - i\csc\theta\partial_\phi)$. We can now determine the Weyl scalar $\psi_{4}$ by first computing the components of the Weyl tensor from the metric components presented in \ref{Newt_metric} and then performing the contraction of the Weyl tensor with the null tetrads, whose components have already been presented above. Hence, the Newtonian limit of $\psi_{4}$ becomes (for a similar expression for $\psi_{0}$, see \cite{LeTiec:2020bos})
\begin{equation}
\lim_{c\rightarrow\infty}c^2\psi_4 =  -\frac{1}{2}\Bar{m}^{i}\Bar{m}^{j}\nabla_{i}\nabla_{j}U~,
\end{equation}
where $\nabla_i$ is a covariant derivative compatible with the three-dimensional Euclidean metric in the Cartesian coordinates. Transforming to the spherical polar coordinates, and using the form for the null tetrad vector $\Bar{m}^{i}$ defined above, along with the form of the Newtonian potential $U$ from \ref{potential_Fourier_space}, we arrive at 
\begin{equation}\label{newtonian_psi4}
\lim_{c\rightarrow\infty}c^2\psi_4=\lim_{c\rightarrow\infty}c^2\sum_{lm}\psi_4^{lm}=\sum_{lm}\alpha_{lm}(t)r^{l-2}\left[1+F_{lm}\left(\frac{R}{r}\right)^{2l+1}\right]\,_{-2}Y_{lm}~,
\end{equation}
where $F_{lm}$ is indeed the tidal response function derived earlier. Note that in arriving at \ref{newtonian_psi4}, connecting the Weyl scalar $\psi_{4}$ and the Newtonian response function $F_{lm}$, we have used the result $\Bar{m}^{i}\Bar{m}^{j}\nabla_{i}\nabla_{j}=(1/2r^{2})\Bar{\eth}_1\Bar{\eth}_0$, where $\Bar{\eth}_s = -(\partial_\theta - i\csc\theta\,\partial_\phi-s\cot\theta)$, with $s$ being an integer and $\Bar{\eth}_s$ being the spin-$s$ lowering operator for spin-weighted spherical harmonics~\cite{Newman:1966ub,LeTiec:2020bos}. Moreover, we have also used the following result: $\Bar{\eth}_1\Bar{\eth}_0Y_{lm}=\sqrt{l(l+1)(l-1)(l+2)}\,_{-2}Y_{lm}$, relating the spherical harmonics with spin-weighted spherical harmonics, to arrive at \ref{newtonian_psi4}. Moreover, the overall time-dependent factor $\alpha_{lm}(t)$ gets related to the tidal field components $\mathcal{E}_{lm}(t)$ through the following relation:
\begin{equation}
\alpha_{lm}(t)=\frac{1}{4}\sqrt{\frac{(l+2)(l+1)}{l(l-1)}}\mathcal{E}_{lm}(t)~.
\end{equation}
Thus, the Weyl scalar $\psi_{4}$ in the Newtonian limit indeed yields the appropriate tidal response function $F_{lm}$ derived earlier through the Newtonian potential and hence provides the way forward for defining a gauge-invariant tidal response function. 
It is worth mentioning that the definition of the tidal response function in terms of the Weyl scalar is indeed relativistically invariant, but it is dependent on the spacetime foliation. This is because the Weyl scalar depends on the choice of the tetrad vectors; any change in the tetrad vectors is going to modify $\psi_{4}$, while for a fixed choice of the tetrad vectors, $\psi_{4}$ does not depend on the choice of the coordinates. However, the response function depends on the asymptotic falloff behavior of $\psi_4$, which for generic coordinate transformations is going to be retained, and in this sense, the tidal response function is gauge invariant. In the subsequent section, we will highlight the key steps in deriving the tidal response function from the Weyl scalar, which in turn will pave the way for defining the tidal Love numbers of black holes under an external tidal field. As we will demonstrate later, there are several issues in the present definition of the tidal Love numbers, when applied to black holes, and this can lead to ambiguities in its definition. 

\subsection{Defining the Love numbers from the tidal response function}\label{def_tln}

Having ascertained the role of $\psi_{4}$ in defining the tidal response function even in the Newtonian limit, let us briefly outline how the response function can be derived in the relativistic setting, and then we will elaborate on extracting the tidal Love numbers from the response function. The starting point is the Teukolsky equation for $\psi_{4}$, which can be solved either numerically or analytically with appropriate approximations. The corresponding solution for $\psi_{4}$, in general, will involve two arbitrary constants of integration. One of them can be fixed by the purely ingoing boundary condition at the black hole horizon (for the corresponding situation in the context of non-black-hole solutions, see \cite{Dey:2020lhq,Dey:2020pth,Chakraborty:2021gdf,Chakraborty:2022zlq,Datta:2020rvo}). Subsequently, after imposing the above boundary condition, one expands the solution for $\psi_{4}$ in a region that is far away from both the black hole and the source of the tidal field, often referred to as the intermediate region, such that the Weyl scalar becomes
\begin{equation}\label{intermediate_psi4}
\psi_4^{\text{intermediate}}=\sum_{lm}\tilde{\alpha}_{lm}(t)r^{l-2}\left[1+F_{lm}\left(\frac{R}{r}\right)^{2l+1}\right]\,_{-2}Y_{lm}~,
\end{equation}
where $\tilde{\alpha}_{lm}(t)$ is a time-dependent quantity (independent of $r$) and $F_{lm}$ defines the tidal response function. As is evident from the above result, in this expansion of $\psi_{4}$ in the intermediate region, the coefficient of the negative power of the radial coordinate is what corresponds to the tidal response function. Note that in the relativistic setting, the contribution of the tidal effects, as well as the response function of the deformed object, can be unambiguously identified by the solution of the Teukolsky equation by promoting $l\in\mathbb{R}$ \cite{LeTiec:2020bos, LeTiec:2020spy}. The above procedure outlines the method of obtaining the tidal response function of a compact object under an external tidal field in a relativistic setting. The determination of the tidal Love numbers from the response function can be achieved in two distinct ways. One of these is consistent with our intuitive understanding that the tidal Love numbers depict the conservative part of the response function, while the other, as we will see, arises from a different perspective and is seemingly plagued with difficulties in interpretation.

The relation between the Love numbers and the tidal response function arises from the connection between the multipole moment and the external tidal field, as expressed in \ref{def_I}, with the tidal Love numbers being the coefficient of the tidal field $\mathcal{E}_{lm}$ (see \cite{poisson_will_2014} for a comprehensive discussion and \cite{Chia:2020yla,Creci:2021rkz,LeTiec:2020bos,LeTiec:2020spy} for recent developments). On the other hand, the coefficient of the time derivative of the tidal field in \ref{def_I} is referred to as tidal dissipation, where the time derivative is taken with respect to the proper time of the observer, corotating with the black hole. Since the multipole moments and the tidal fields are real (these are observables in classical physics), it follows that the tidal Love numbers $k_{lm}$ and the dissipation term $\tau_{0}\nu_{lm}$ must also be real. Thus, the real part of the response function, at least in the small-frequency approximation, must correspond to the tidal Love numbers. On the other hand, the imaginary part of the response function, possibly proportional to the frequency in the corotating frame of reference, should describe the dissipative effects. This is also consistent with our intuition that the tidal Love numbers correspond to the conservative part of the response function and hence must be real, while the tidal dissipation, as the name suggests, is related to the dissipative part and thus must be imaginary. Along these lines, we may argue that the tidal Love numbers are simply the real part of the response function, modulo a factor of (1/2), while the imaginary part of the response function describes the tidal dissipation. With this definition we obtain, in the small-frequency limit, the following result for the Love numbers:
\begin{equation}\label{Love_N_1}
k^{\rm (1)}_{lm}\equiv \frac{1}{2}\textrm{Re}F_{lm}~,
\end{equation}
while the dissipative part is given by 
\begin{align}\label{dissip_1}
\omega' \tau_{0}\nu^{\rm (1)}_{lm}\equiv \textrm{Im}F_{lm}~.
\end{align}
To reiterate, the fact that Love numbers are the real part of the tidal response function is consistent with the expectation that Love numbers are conservative in nature, while by definition the imaginary part must be dissipative. As we will see, for the Schwarzschild black hole, the dissipative part is independent of the frequency and depends only on the mass of the black hole. On the other hand, for the slowly rotating Kerr black hole, the dissipative part will become frequency dependent, with a finite zero-frequency limit. We advocate this proposal for the determination of the tidal Love numbers and the dissipative part, from the response function, most strongly, since it neatly connects the expectations from the earlier results in the literature to the intuitive picture and can be straightforwardly generalized to determine the Love numbers of an arbitrary rotating Kerr black hole.  

Apart from the definitions of the tidal Love numbers and the dissipative part given in \ref{Love_N_1} and \ref{dissip_1}, one can also define the tidal Love numbers from the tidal response function in the following manner:
\begin{equation}\label{Love_N_2}
k^{\rm (2)}_{lm}\equiv \frac{1}{2}\times\textrm{term~independent~of~} \omega' \textrm{~in~} F_{lm}~,
\end{equation}
with the dissipative part  
\begin{align}\label{dissip_2}
\omega' \tau_{0}\nu^{\rm (2)}_{lm}\equiv \textrm{term~dependent~of~} \omega' \textrm{~in~} F_{lm}~.
\end{align}
This method is motivated by \ref{def_I}, and uses the fact that in the frequency domain, the dissipative part has an overall factor of $i\omega'$, where $\omega'$ is the frequency in the corotating frame of reference. In the case of a nonrotating black hole, the dissipative part becomes proportional to $i\omega$.

Note that, for nonrotating black holes, we simply have to replace $\omega'$ with $\omega$ in the above definition. We would like to emphasize that the above definition of the tidal Love numbers is motivated by the analysis of \cite{poisson_will_2014}, which is performed in a nonrelativistic and weak field regime. Therefore, it is worthwhile to ask if one would expect the above formalism to hold true in the context of an arbitrarily rotating black hole and in particular, for an extremal black hole. In this respect, the above approach is questionable and is also counterintuitive, as the tidal Love numbers can become imaginary, and hence can also induce dissipative effects. Though there are claims in the literature that the imaginary part of the tidal Love numbers can be due to a phase lag between the tidal field and the deformation in the compact object~\cite{LeTiec:2020bos,LeTiec:2020spy}, it is not clear why this lag should persist even in the time-independent situation, as any such lag should dissipate out with time.

From our point of view, the first approach, where the tidal Love numbers are related to the real part of the response function seems the appropriate one. This is also due to the fact that the Love numbers themselves can be frequency dependent, but if we define the Love numbers as the leading-order term in an expansion in powers of $M\omega$, they cannot possibly encapsulate the same. In this respect as well, we can consider the Love numbers to be simply the conservative part, which can be determined from the real part of the tidal response function. Nonetheless, in what follows, we will present the expressions of the tidal Love numbers and the dissipative part explicitly, in both of these approaches.

\section{Teukolsky equation in the small-frequency limit}\label{small_freq_master}

As demonstrated in the earlier sections, the tidal response function can be defined in a relativistic manner through the asymptotic expansion of the perturbed Weyl scalar $\psi_{4}$. These perturbations of the Weyl scalar satisfy the Teukolsky equation, which arises from the gravitational perturbation of a perturbed Kerr black hole \cite{Teukolsky:1972my, Teukolsky:1973ha, Press:1973zz, Teukolsky:1974yv}. Thus, from the solution of the Teukolsky equation, one can extract the tidal response function through an asymptotic expansion, which has been attempted recently in \cite{LeTiec:2020spy, LeTiec:2020bos, Chia:2020yla}. From the tidal response function, derived through the solution of the Teukolsky equation, the tidal Love numbers of both rotating and nonrotating black holes have been derived. Though \cite{LeTiec:2020spy, LeTiec:2020bos} predict nonzero but imaginary tidal Love numbers, \cite{Chia:2020yla} argues that the nonzero part of the response function is arising from dissipative effects, while the Love numbers, which depict the conservative part, are actually zero. To show the same, it is important to solve the Teukolsky equation in the small-frequency approximation ($M\omega\ll1$) as in \cite{Chia:2020yla}, rather than simply considering the static limit \cite{LeTiec:2020spy, LeTiec:2020bos}. However, in the analysis of \cite{Chia:2020yla}, several linear-order terms in $M\omega$, appearing in the Teukolsky equation, were neglected. Therefore, the result for the tidal response function, as quoted in \cite{Chia:2020yla}, in terms of the mode frequency $\omega$, is at best incomplete. 

In this work, we will take all those terms into account, which had been neglected in the approximated low-frequency Teukolsky equation of \cite{Chia:2020yla}, and then present the solution of the Teukolsky equation considering all the linear-order terms in $M\omega$. Subsequently, we will present the black hole's response function and the tidal Love numbers. The main differences between our work and reference \cite{Chia:2020yla} are as follows:
\begin{itemize}

    \item In the analysis of \cite{Chia:2020yla}, the Teukolsky equation in the small-frequency limit has been solved, but some linear-order terms in $M\omega$ have been neglected, which in principle should be present. In this work, we have included all those terms in the Teukolsky equation and hence determine the corresponding solution accurately up to linear order in $M\omega$. The resulting expression for the tidal response function is different from the one derived in \cite{Chia:2020yla}. 
    
    \item The calculation of the response function, as presented in \cite{Chia:2020yla}, is based on the small $M\omega$ approximation. However, the decomposition of the response function into the tidal Love numbers and tidal dissipation for a rotating compact object is based on the $M\omega'$ expansion, where $\omega' = \omega - m\Omega_{\rm h}$ is the frequency observed in the frame, corotating with the black hole. As is evident, for arbitrary rotation, small values of $M\omega$ do not imply that $M\omega'$ will be small as well unless the black hole is nonrotating or slowly rotating.
    Thus, the  analysis of \cite{Chia:2020yla} is only applicable to nonrotating and slowly rotating black holes. In this work, we have elaborated on this result and have also proposed a possible generalization of the decomposition of the tidal response function for arbitrarily large rotations. 
    
\end{itemize}

Having outlined the shortcomings of \cite{Chia:2020yla}, we provide below the master equation satisfied by the radial part of the gravitational perturbation. For this purpose, we employ Teukolsky's formulation of black hole perturbation~\cite{Teukolsky:1972my, Teukolsky:1973ha, Press:1973zz, Teukolsky:1974yv}, based on the Newman-Penrose formalism~\cite{Newman:1961qr}. In general, Teukolsky's formalism provides the master equation for various perturbations, each characterized by a single spin value (see \cite{Teukolsky:1973ha}). For example, the equation for $s=0$ depicts the evolution of a test scalar field, $s=(1/2)$ describes a test neutrino field, $s=\pm 1$ presents a test electromagnetic field, and finally, $s=\pm 2$ corresponds to  gravitational perturbations. For our purpose, we are interested in the gravitational perturbation described by the perturbed Weyl scalar $\psi_{4}$; therefore, we will consider the Teukolsky equation with $s=-2$. Even though the Teukolsky equation is often quoted in the Boyer-Lindquist coordinate system $(t,r,\theta,\phi)$, for our purpose it will be convenient if we express the Teukolsky equation in the ingoing Kerr coordinates $(v,r,\theta,\Tilde{\phi})$. These ingoing Kerr coordinates are related to the Boyer-Lindquist coordinates through the following transformations~\cite{Teukolsky:1974yv}:
\begin{equation}
\mathrm{d}v = \mathrm{d}t+\frac{r^2+a^2}{\Delta}\mathrm{d}r, \qquad \mathrm{d}\Tilde{\phi} = \mathrm{d}\phi +\frac{a}{\Delta}\mathrm{d}r~,
\end{equation}
such that the Kerr metric in the ingoing Kerr coordinates becomes \cite{Chia:2020yla, Teukolsky:1974yv}
\begin{multline}
\mathrm{d}s^2 = -\left(1-\frac{2Mr}{\Sigma}\right)\mathrm{d}v^2 + 2\mathrm{d}v\mathrm{d}r-\frac{4Mar\sin^2\theta}{\Sigma}\mathrm{d}v\mathrm{d}\Tilde{\phi}-2a\sin^2\theta\mathrm{d}r\mathrm{d}\Tilde{\phi}
\\
+\Sigma\mathrm{d}\theta^2+\frac{\left[(r^2+a^2)^2-a^2\Delta\sin^2\theta\right]\sin^2\theta}{\Sigma}\mathrm{d}\Tilde{\phi}^2~,
\end{multline}
where $\Sigma = r^2+a^2\cos^2\theta$, and $\Delta = r^2-2Mr+a^2$. Given the Kerr metric in the ingoing Kerr coordinates, we will express both the Weyl scalar as well as the associated Teukolsky equation in this system of coordinates.

First of all, we express the perturbed Weyl scalar $\psi_{4}$ in the ingoing Kerr coordinates, decomposed in the radial and the angular parts as \cite{Teukolsky:1974yv} 
\begin{equation}
\rho^4 \psi_4 = \int \mathrm{d}\omega\,e^{-i\omega v}\sum_{lm} e^{-im\Tilde{\phi}}\,_{-2}S_{lm}(\theta)R(r)~,
\end{equation}
where $\rho = -(r-ia\cos\theta)$, the angular part $_{-2}S_{lm}$ is the spin-weighted spheroidal harmonic, and the Weyl scalar has been constructed using the Kinnersley tetrad~\cite{Kinnersley:1969zza}. 
Finally, the radial perturbation $R(r)$ satisfies the following equation in the ingoing Kerr coordinates~\cite{Chia:2020yla}:
\begin{multline}
\frac{\mathrm{d}^2R}{\mathrm{d}r^2} + \left(\frac{2iP_+-1}{r-r_+}-\frac{2iP_-+1}{r-r_-}-2i\omega\right)\frac{\mathrm{d}R}{\mathrm{d}r} \\ +\left\{\frac{4iP_-}{(r-r_-)^2}-\frac{4iP_+}{(r-r_+)^2}+\frac{A_-+iB_-}{(r-r_-)(r_+-r_-)}-\frac{A_++iB_+}{(r-r_+)(r_+-r_-)}\right\}R=\frac{T}{\Delta}~,
\label{chiaA41}
\end{multline}
where $r_\pm = M\pm\sqrt{M^2-a^2}$ are the event and the Cauchy horizons, respectively. Further, the other constants appearing in the above differential equation read \cite{Chia:2020yla, Press:1973zz} 
\begin{align}
P_\pm &= \frac{am-2r_\pm M\omega}{r_+-r_-},\,\,\,\,\,\,B_\pm = 2r_\pm\omega~,
\nonumber
\\
A_\pm &= E_{lm}-2-2(r_+-r_-)P_\pm\omega-(r_\pm+2M)r_\pm\omega^2~,
\nonumber
\\
E_{lm}&=  l(l+1) -2a\omega\frac{4m}{l(l+1)} + O[(a\omega)^2]~.
\end{align}
Finally, using the following transformation from the radial coordinate $r$ to the rescaled dimensionless coordinate $z$, defined as, $z=(r-r_+)/(r_+-r_-)$, the radial Teukolsky equation becomes (for a derivation, see \ref{app_A})
\begin{multline}
\frac{\mathrm{d}^2R}{\mathrm{d}z^2} + \left\{\frac{2iP_+-1}{z} -\frac{2iP_1+1}{z+1}\right\}\frac{\mathrm{d}R}{\mathrm{d}z} + \left[\frac{4iP_2}{(z+1)^2}-\frac{4iP_+}{z^2}-\frac{l(l+1) -2}{z(1+z)} \right.\\+\left.\frac{2ma\omega}{z(1+z)}\left\{1+\frac{4}{l(l+1)}\right\} -\frac{2i\omega r_+}{z(z+1)} \right]R=0\,,
\label{app_gen_eq}
\end{multline}
where $P_1 = P_- + \omega (r_+-r_-)$ and $P_2 = P_{-}-(1/2)\omega (r_+-r_-)$. In the above, we have ignored terms $\mathcal{O}(\omega^{2})$, but we have kept all the terms $\mathcal{O}(M\omega)$, unlike \cite{Chia:2020yla}, where several terms $\mathcal{O}(M\omega)$ had been ignored. We will now solve the above master equation, and from the asymptotic behavior of the solutions, we will determine the tidal response function for nonrotating and slowly rotating black holes in the small-frequency limit. We start by computing the tidal response function of the Schwarzschild black hole in the subsequent section.

\section{Tidal response of a Schwarzschild black hole}\label{response_Schwarzschild}

In this section, we present the tidal response of a Schwarzschild black hole under the influence of an external tidal field. For this purpose, we start with the radial perturbation equation derived in the previous section, keeping all terms $\mathcal{O}(M\omega)$, and then substitute the rotation parameter $a=0$. This yields the following differential equation for the radial part of the perturbation:
\begin{equation}
\frac{\mathrm{d}^2R}{\mathrm{d}z^2} + \left\{\frac{2iP_+-1}{z} +\frac{2iP_+-1}{z+1}\right\}\frac{\mathrm{d}R}{\mathrm{d}z} + \left\{\frac{2iP_+}{(z+1)^2}-\frac{4iP_+}{z^2}-\frac{l(l+1) -2}{z(1+z)} +\frac{2iP_+}{z(z+1)} \right\}R=0~.
\end{equation}
In order to arrive at the above equation, we also used the fact that in the limit of zero rotation, one has $r_{+}=2M$ and $r_{-}=0$, and consequently,
$P_{-}=0$, $P_{+}=-2M\omega=-P_{1}$, and $P_{2}=(P_{+}/2)$. The above equation, for zero rotation, can be solved exactly and the solutions can be written in terms of hypergeometric functions as follows:
\begin{multline}
R(z) =  (1+z)^{2-3iP_+}\Big[z^{-2 i P_+} C_2 \,_{2}F_{1}\left(-l -3iP_+\frac{2l-1}{2l+1}, 
     1+l -3iP_+\frac{2l+3}{2l+1}; -1 - 2 i P_+; -z\right) 
     \\\left.
   +z^{2}  C_1 \,_{2}F_{1}\left(2-l -iP_+\frac{2l-5}{2l+1}, 3+l -iP_+\frac{2l+7}{2l+1}; 
     3 + 2 i P_+; -z\right)\right],
\end{multline}
where $C_1$ and $C_2$ are the two arbitrary constants of integration. All arguments of both hypergeometric functions as well as all other terms in the above solution, including the power of $(1+z)$, are expressed up to linear order in $M\omega$. Using a useful property of the hypergeometric function namely, $\,_{2}F_{1}(a,b;c;z\rightarrow 0)=1$~\cite{abramowitz_stegun_1972}---it follows that in the near-horizon regime the radial perturbation $R(z)$ becomes 
\begin{equation}
R(z)\simeq z^{-2 i P_+} C_2 +z^{2} C_1~.
\end{equation}
The radial function associated with $C_{2}$ simplifies as $z^{-2iP_{+}}=e^{4iM\omega \ln z}$, which corresponds to an outgoing mode at the horizon, and hence must be absent in the black hole spacetime. This physical expectation implies that
$C_2$ must vanish.
Therefore, by imposing the ingoing boundary condition at the horizon, the radial part of gravitational perturbation becomes 
\begin{equation}\label{radial_schwarzschild}
R(z) =  (1+z)^{2-3iP_+}z^{2}  C_1 \,_{2}F_{1}\left(2-l -iP_+\frac{2l-5}{2l+1}, 3+l -iP_+\frac{2l+7}{2l+1}; 3 + 2 i P_+; -z\right)~.
\end{equation}
It is evident that this solution differs from the one given in Ref. \cite{Chia:2020yla} by terms $\mathcal{O}(M\omega)$, as expected. Given the profile for the radial perturbation in the near zone, one can determine the radial part of the perturbed Weyl scalar $\psi_4$, which is simply given by $\psi_4 \propto (1+z)^{-4} R(z)$. Thus, using \ref{radial_schwarzschild}, we obtain the following radial dependence for $\psi_{4}$:
\begin{equation}\label{weyl_radial_Sch}
\psi_4 \propto  z^{2} (1+z)^{-2-Q_3}\,_{2}F_{1}\left(2-l-Q_1, 3+l -Q_2; 3 + 2 i P_+; -z\right)~,
\end{equation}
where we have defined the following quantities:
\begin{equation}\label{Q_exp}
Q_1 = -2iM\omega\frac{2l-5}{2l+1},\,\,\,\,\,Q_2 = -2iM\omega\frac{2l+7}{2l+1},\,\,\,\,Q_3=-6iM\omega~.
\end{equation}
Note that each of these quantities is proportional to $M\omega$, which is assumed to be small, in our calculations. It is also worth emphasizing that each of these terms---namely, $Q_{1}$, $Q_{2}$, and $Q_{3}$---are absent in the corresponding expression of the Weyl scalar $\psi_{4}$ in \cite{Chia:2020yla}, and thus, it is expected that the tidal response function derived here will be different. In particular, it will be interesting if the modified tidal response function predicts nonzero tidal Love numbers for the Schwarzschild black hole. 

The determination of the tidal response of the Schwarzschild black hole to an external tidal field can be obtained by considering the large-$r$ limit of the perturbed Weyl scalar $\psi_{4}$. Using the asymptotic expansion of the hypergeometric function, \ref{weyl_radial_Sch} yields
\begin{multline}\label{asymp_sch_weyl}
\psi_4(r\rightarrow \infty) \propto z^{l -2 + Q_1 - Q_3} \frac{\Gamma\left(3 + 2 i P_+\right) \Gamma\left(1 + 2 l + Q_1 - Q_2\right)}{\Gamma\left(1 + l + 2 i P_+ + Q_1\right) \Gamma\left(3 + l - Q_2\right)}
\\
\times\left[1+ z^{-2l-1 + Q_2-Q_1} \frac{\Gamma\left(1 + l + 2 i P_+ + Q_1\right) \Gamma\left(3 + l - Q_2\right) \Gamma\left(-1 - 2 l - Q_1 + Q_2\right)}{ \Gamma\left(1 + 2 l + Q_1 - Q_2\right)\Gamma\left(2 - l - Q_1\right) \Gamma\left(-l + 2 i P_+ + Q_2\right)}\right]~.
\end{multline}
Since $Q_1$, $Q_2$, and $Q_3$ are small under the approximations for which the above computation holds, we can directly compare the above expression for $\psi_{4}$ with that in \ref{intermediate_psi4}. This yields the following response function for a Schwarzschild black hole under the influence of an external tidal field:
\begin{equation}
F_{\text{Sch}} = \frac{\Gamma\left(1 + l + 2 i P_+ + Q_1\right) \Gamma\left(3 + l - Q_2\right) \Gamma\left(-1 - 2 l - Q_1 + Q_2\right)}{ \Gamma\left(1 + 2 l + Q_1 - Q_2\right)\Gamma\left(2 - l - Q_1\right) \Gamma\left(-l + 2 i P_+ + Q_2\right)}\,.
\label{resp_Sch}
\end{equation}
Note that, on ignoring the quantities $Q_{1}$ and $Q_{2}$, the tidal response function, as given above, reduces to the one derived in \cite{Chia:2020yla}, as expected. Thus, indeed, the response function gets modified on the inclusion of all the terms of $\mathcal{O}(M\omega)$. This can be observed more explicitly by expanding out these gamma functions to leading order in $M\omega$, and keeping in mind that for gravitational perturbations $l\in\mathbb{L}=\mathbb{Z}^+\backslash\{1\}$, yielding (for detailed calculation, see \ref{app_B})
\begin{equation}
F_{\text{Sch}}=-\frac{1}{2}\frac{\Gamma\left(1 + l\right) \Gamma\left(3 + l\right) \Gamma\left(l-1\right)\Gamma\left(1+l\right)}{\Gamma\left(1 + 2 l\right)\Gamma\left(2l+2\right)}\left[2 i P_++\frac{1}{2}( Q_2 - Q_1)\right]~.
\label{resp_Sch_1}
\end{equation}
For a Schwarzschild black hole, it follows that $P_{+}=-2M\omega$, and from \ref{Q_exp}, the difference $Q_{2}-Q_{1}$ becomes $-2iM\omega\{12/(2l+1)\}$. Therefore, the tidal response function for the Schwarzschild black hole can be reexpressed as 
\begin{equation}\label{sch_response_final}
F_{\text{Sch}}=2i M\omega\frac{\left(2+l\right)! \left(l-2\right)!\left(l!\right)^{2}}{\left(2 l\right)!\left(2l+1\right)!}\left(1+\frac{3}{2l+1}\right)~.
\end{equation}
As evident, the term outside the round bracket is the response function derived in \cite{Chia:2020yla}, provided terms $\mathcal{O}(M^{2}\omega^{2})$ have been neglected, while the term inside the round bracket corresponds to the correction arising due to keeping additional terms of $\mathcal{O}(M\omega)$ in our analysis. Intriguingly, despite the inclusion of all the correction terms of $\mathcal{O}(M\omega)$, the response function of the Schwarzschild black hole is purely imaginary and the imaginary part is proportional to $M\omega$. Therefore, according to \ref{Love_N_1} and \ref{Love_N_2}, it follows that the Schwarzschild black hole has zero tidal Love numbers---in particular, 
\begin{equation}
k_{lm}^{(1)}=0=k_{lm}^{(2)}~.
\end{equation}
In words, the tidal Love numbers, derived using both of the approaches, identically vanish. Along similar lines, the dissipative parts obtained using both of these approaches, as outlined in \ref{dissip_1} and \ref{dissip_2}, also coincide. 
\begin{equation}\label{sch_tid_dis}
\nu_{lm}^{(1)}=\frac{2M}{\tau_{0}}\frac{\left(2+l\right)! \left(l-2\right)!\left(l!\right)^{2}}{\left(2 l\right)!\left(2l+1\right)!}\left(1+\frac{3}{2l+1}\right)=\nu_{lm}^{(2)}~,
\end{equation}
however, is different from \cite{Chia:2020yla}. To make the difference explicit, we have quoted numerical values of the response function from \ref{sch_response_final}, for different choices of the angular number $l$ in \ref{t1} and have compared them with the corresponding values derived from the response function of \cite{Chia:2020yla}. As is evident, the coefficient of $M\omega$ is different in the two cases, and hence the tidal dissipation, whose correct expression is given by \ref{sch_tid_dis}, is larger compared to the corresponding expression in \cite{Chia:2020yla}. This is also clear from \ref{t1}. 
\begin{table}[ht]
    \centering
    \setlength{\tabcolsep}{20pt}
    \renewcommand{\arraystretch}{1.5}
    \begin{tabular}{lll}
    \hline
        $l$ & Response function of \cite{Chia:2020yla} & Response function of this work \\ \hline
        2 & $(0 + 0.0666667 \,i) \,M\omega$ & $(0 + 0.106667 \,i) \,M\omega$ \\ 
        3 & $(0 + 0.00238095 \,i) \,M\omega$ & $(0 + 0.00340136 \,i) \,M\omega$ \\ 
        4 & $(0 + 0.000113379 \,i) \,M\omega$ & $(0 + 0.000151172 \,i) \,M\omega$ \\ \hline
    \end{tabular}
\caption{Numerical values of the tidal response function computed in this work for Schwarzschild black hole of mass $M$ has been presented, and contrasted with those obtained in Ref.~\cite{Chia:2020yla}. As is evident, the tidal Love numbers vanish in both the cases, while our result, which incorporates all the terms $\mathcal{O}(M\omega)$, predicts a larger tidal dissipation in comparison with the corresponding term in \cite{Chia:2020yla}.}
\label{t1}
\end{table}

\begin{figure}[ht]
    \centering
    \includegraphics[scale=0.8]{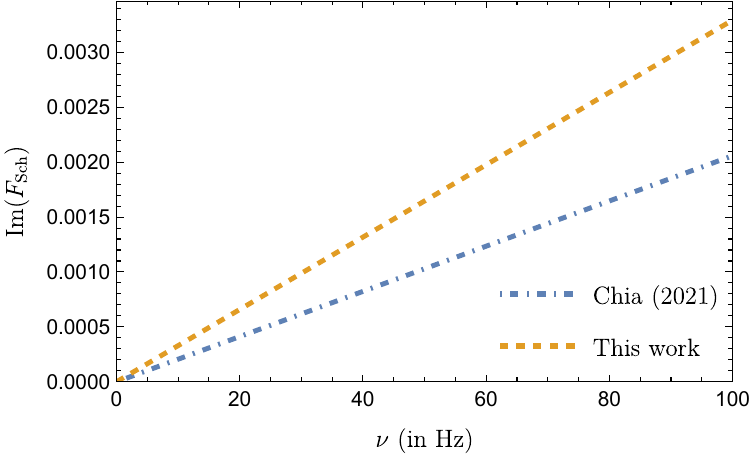}
    \caption{The imaginary part of the tidal response function $F_{\rm Sch}$ has been plotted against the frequency $\nu$ (in Hz) for a Schwarzschild black hole of mass $10\,M_{\odot}$, and the corresponding result derived in the present work has been compared with the corresponding result in \cite{Chia:2020yla}. See text for more discussions.}
    \label{10Ms_nr_plot}
\end{figure}

To further bolster our claim, we have plotted the variation of the imaginary part of the response function $F_{\rm Sch}$, describing the part corresponding to tidal dissipation, against the frequency $\nu\equiv(\omega/2\pi)$, expressed in standard units, in \ref{10Ms_nr_plot} for the $l=m=2$---i.e., the dominant gravitational wave mode. As expected, the plots of the tidal dissipation are linear in frequency, while the slopes of the curves are different when the result of the present work is compared with that of \cite{Chia:2020yla}. Moreover, for a given frequency, it follows that the actual value of the tidal dissipation, as derived in the present work, is large compared to that of \cite{Chia:2020yla}. This is expected, as the actual response function derived here is the response function in \cite{Chia:2020yla} multiplied by a quantity larger than unity [see, e.g., \ref{sch_response_final}]. Thus, inclusion of all the terms of $\mathcal{O}(M\omega)$ appearing in the Teukolsky equation, leads to a more complete expression for the tidal response function and differs from \cite{Chia:2020yla}. In the next section, we will explicitly demonstrate that the same is true for a slowly rotating Kerr black hole as well.  

\section{Tidal response of a slowly rotating Kerr black hole}\label{response_slow}

In the previous section, we have explicitly demonstrated that the tidal response function of a Schwarzschild black hole differs from the one derived in \cite{Chia:2020yla}, due to the inclusion of several terms, all of which are $\mathcal{O}(M\omega)$. It is intriguing that even after all these modifications to the tidal response function, the tidal Love numbers associated with the Schwarzschild black hole identically vanish. However, the tidal dissipation is different from \cite{Chia:2020yla} by terms of $\mathcal{O}(M\omega)$. In this section, we wish to extend our analysis to slowly rotating black holes---i.e., we will work up to the linear order in $(a/M)$ and shall neglect all the higher-order terms $\mathcal{O}(a^{2}/M^{2})$. This is because the tidal response function is expanded in terms of frequencies $\omega'=\omega-m\Omega_{h}$ in the corotating frame of reference, while the small-frequency expansion of the Teukolsky equation involves ignoring terms $\mathcal{O}(M^{2}\omega^{2})$. These two expansions, in general, are very different---small $M\omega$ does not necessarily mean small $M\omega'$---except when rotation is small. Thus, the decomposition of the tidal response function into the conservative Love numbers and the dissipative part, as defined in \cite{Chia:2020yla}, is only applicable for small rotation parameters. We will discuss various possibilities of extending the corresponding decomposition of the tidal Love numbers for generic rotation in the subsequent sections.   

In the limit of slow rotation ($a/M\ll1$), the master equation for the radial perturbation---namely, \ref{app_gen_eq}---becomes
\begin{multline}
\frac{\mathrm{d}^2R}{\mathrm{d}z^2} + \left\{\frac{2iP_+-1}{z} -\frac{2iP_++1+8iM\omega}{z+1}\right\}\frac{\mathrm{d}R}{\mathrm{d}z} + \left[\frac{4i(P_+ + M\omega)}{(z+1)^2}-\frac{4iP_+}{z^2}-\frac{l(l+1) -2}{z(1+z)}\right.
\\
\left.+\frac{2ma\omega}{z(1+z)}\left\{1+\frac{4}{l(l+1)}\right\} -\frac{4iM\omega}{z(z+1)} \right]R=0~.
\end{multline}
Here, we have used the result that in the limit of slow rotation, $r_{-}=\mathcal{O}(a^{2})$, $r_{+}=2M+\mathcal{O}(a^{2})$, such that, $P_{+}=\{(am-2M\omega r_{+})/r_{+}\}+\mathcal{O}(a^{2})$ and $P_{-}=(am/r_{+})+\mathcal{O}(a^{2})$. It turns out that the above differential equation can also be solved exactly in terms of the hypergeometric functions, with the following solution: 
\begin{multline}
R(z) =   C_1\, z^{-2 i P_+} (1 + z)^{2 + 6 i M \omega}
\\ 
\times\,_{2}F_{1}\left(-l - 2 i P_+ + 2 i M \omega\frac{2 l-5}{2 l+1} , 1 + l - 2 i P_+ + 2 i M \omega\frac{2 l+7}{2 l+1}; -1 - 2 i P_+; -z\right) 
\\ 
+C_2 \,z^{2} (1 + z)^{2 + 6 i M \omega}\,_{2}F_{1}\left(2 - l + 2 i M \omega\frac{2 l-5}{2 l+1}, 3 + l + 2 i M \omega\frac{2 l+7}{2 l+1}; 3 + 2 i P_+; -z\right)~,
\end{multline}
where $C_1$ and $C_2$ are again the arbitrary constants of integration, and all the terms in the above solution, including the arguments of the hypergeometric functions, are written up to linear orders of $M\omega$ and $(a/M)$. All the second- and higher-order terms of $M\omega$ and $(a/M)$ have been neglected, along with the terms involving the multiplication of $M\omega$ and $(a/M)$. For small $z$---i.e., in the near-horizon limit---the hypergeometric functions become unity, and the term associated with the arbitrary constant $C_{1}$ behaves as $e^{i\omega'r_{*}}$, representing an outgoing wave. Since we are interested in black hole geometries, it follows that outgoing modes should be absent and hence we must set $C_{1}=0$. Therefore, by imposing the ingoing boundary condition at the horizon, the solution for the radial perturbation equation simplifies to 
\begin{equation}
R(z) =  C_2 \,z^{2} (1 + z)^{2 + 6 i M \omega}\,_{2}F_{1}\left(2 - l + 2 i M \omega\frac{2 l-5}{2 l+1}, 3 + l + 2 i M \omega\frac{2 l+7}{2 l+1}; 3 + 2 i P_+; -z\right)~.
\end{equation}
Given the radial function $R(z)$, the radial part of the perturbed Weyl scalar $\psi_4$ can be obtained by noting that in the near-horizon regime, $\psi_4$ is approximately proportional to $(1+z)^{-4}R(z)$. Therefore the radial part of $\psi_{4}$ becomes
\begin{equation}
\psi_4 \propto  z^{2} (1+z)^{-2-Q_3}\,_{2}F_{1}\left(2-l -Q_1, 3+l -Q_2; 3 + 2 i P_+; -z\right)~,
\end{equation}
where the quantities $Q_{1}$, $Q_{2}$, and $Q_{3}$ have been defined in \ref{Q_exp} and are all $\mathcal{O}(M\omega)$. Note that these are precisely the terms absent in \cite{Chia:2020yla} and are going to affect the tidal response function. 

The computation of the tidal response function of a slowly rotating black hole under an external tidal field can be obtained by taking the large-$r$ limit of the perturbed Weyl scalar $\psi_{4}$. The expansion of the hypergeometric function for its large argument involves a growing term and a decaying term. The growing term corresponds to the external tidal field, and the decaying term relates to the response of the body under it. Thus, in the far-zone region, the Weyl scalar $\psi_{4}$ reads
\begin{multline}
    \psi_4 \propto z^{l -2 + Q_1 - Q_3} \frac{\Gamma\left(3 + 2 i P_+\right) \Gamma\left(1 + 2 l + Q_1 - Q_2\right)}{
  \Gamma\left(1 + l + 2 i P_+ + Q_1\right) \Gamma\left(3 + l - Q_2\right)}\\\times
  \left[1+ z^{-2l-1 + Q_2-Q_1} \frac{\Gamma\left(1 + l + 2 i P_+ + Q_1\right) \Gamma\left(3 + l - Q_2\right) \Gamma\left(-1 - 2 l - Q_1 + Q_2\right)}{ \Gamma\left(1 + 2 l + Q_1 - Q_2\right)
  \Gamma\left(2 - l - Q_1\right) \Gamma\left(-l + 2 i P_+ + Q_2\right)}\right]~.
\end{multline}
Since all of the terms involving $Q_1$, $Q_2$, and $Q_3$ are small, we can immediately compare the above expression for $\psi_{4}$ with \ref{intermediate_psi4}, such that the tidal response function for a slowly rotating Kerr black hole becomes
\begin{equation}
    F_{\text{Kerr(slow)}} = \frac{\Gamma\left(1 + l + 2 i P_+ + Q_1\right) \Gamma\left(3 + l - Q_2\right) \Gamma\left(-1 - 2 l - Q_1 + Q_2\right)}{ \Gamma\left(1 + 2 l + Q_1 - Q_2\right)
  \Gamma\left(2 - l - Q_1\right) \Gamma\left(-l + 2 i P_+ + Q_2\right)}~.
  \label{resp_Kerr}
\end{equation}
By expanding the response function to the linear order in $M\omega$, and for $l\in\mathbb{L}=\mathbb{Z}^+\backslash\{1\}$ (for a derivation, see \ref{app_B}),
\begin{equation}
    F_{\text{Kerr(slow)}} =
    -\frac{1}{2}\frac{\Gamma\left(1 + l\right) \Gamma\left(3 + l\right) \Gamma\left(l-1\right)\Gamma\left(1+l\right)}{\Gamma\left(1 + 2 l\right)\Gamma\left(2l+2\right)}\left[2 i P_{+}+\frac{1}{2}( Q_2 - Q_1)\right]~,
    \label{resp_Kerr_1}
\end{equation}
where second- and higher-order terms of $M\omega$ and $(a/M)$ have been neglected. Intriguingly, the formal structure of the above response function is identical to that of the Schwarzschild black hole; however, the quantity $P_{+}$ is different in the Schwarzschild and the Kerr spacetimes. Hence the tidal response functions will also be different. In particular, writing down the $P_{+}$ explicitly, we obtain the following structure for the tidal response function:
\begin{equation}
    F_{\text{Kerr(slow)}} =
    -\frac{\left(2+l\right)! \left(l-2\right)!\left(l!\right)^{2}}{\left(2 l\right)!\left(2l+1\right)!}\left[i \frac{am}{2M} -2iM\omega\left\{1+\frac{3}{2l+1}\right\}\right]~.
\end{equation}
Note that for $a=0$, it reduces to \ref{sch_response_final}, describing the response function of the Schwarzschild black hole, as it should. Also, in the limit of zero frequency, for $l=2$, the response function becomes $F_{\rm Kerr}=-(ima/60M)$, which coincides with \cite{LeTiec:2020bos,LeTiec:2020spy}. This shows that the above expression for the tidal response function reproduces known results in the literature. Interestingly, in the zero-frequency limit, none of these correction terms ($Q_{1}$, $Q_{2}$, and $Q_{3}$) contributes, and hence, the zero-frequency results will not change by the corrections derived in this work. 

If we consider the tidal Love numbers to be the real part of the response function, which is motivated by the fact that the tidal Love numbers are the conservative part and hence must be real, then the tidal Love numbers of the slowly rotating Kerr black hole vanish identically. However, if we follow the definition of the tidal Love numbers as in \ref{def_resp_sr}, then we must express the response function in terms of the corotating frequency $\omega'=\omega - m\Omega_h$, where $\Omega_h = a/(2Mr_+)$ is the angular velocity of the horizon~\cite{Teukolsky:1974yv}. This yields 

\begin{equation}\label{slow_fkerr}
    F_{\text{Kerr(slow)}} =
    \frac{\left(2+l\right)! \left(l-2\right)!\left(l!\right)^{2}}{\left(2 l\right)!\left(2l+1\right)!}\left[\frac{6imM\Omega_h}{2l+1}+2iM\omega'\left\{1+\frac{3}{2l+1}\right\}\right]~.
\end{equation}
Note that the first term inside the square bracket, as well as the second term inside the curly bracket, is absent in \cite{Chia:2020yla}. Therefore, according to both \ref{Love_N_1} and \ref{Love_N_2}, the tidal Love numbers of the slowly rotating Kerr black hole, as derived from the response function of \cite{Chia:2020yla}, would vanish. However, when all the terms of $\mathcal{O}(M\omega)$ are included in the analysis, as in the present work, it follows that \ref{Love_N_1} yields vanishing Love numbers, while \ref{Love_N_2} predicts nonzero, but imaginary tidal Love numbers, such that
\begin{equation}
k^{(1)}_{\textrm{Kerr(slow)}\,lm}=0~;
\qquad 
k^{(2)}_{\textrm{Kerr(slow)}\,lm}=\frac{1}{2}\frac{\left(2+l\right)! \left(l-2\right)!\left(l!\right)^{2}}{\left(2 l\right)!\left(2l+1\right)!}\left(\frac{6imM\Omega_h}{2l+1}\right)~,
\end{equation}
which is proportional to the angular velocity of the horizon. Thus if we define the tidal response function in the corotating frame of reference, and the tidal Love numbers are the part independent of the corotating frequency $\omega'$, then the tidal Love numbers of a slowly rotating Kerr black hole are nonzero and imaginary. Though the imaginary value is difficult to explain, following \cite{LeTiec:2020bos,LeTiec:2020spy}, one may argue the existence of the imaginary value to be due to a tidal lag effect. On the other hand, this will predict an imaginary quadrupole moment for the slowly rotating Kerr black hole, which is counterintuitive. The dissipative part, on the other hand, as obtained by both the methods outlined in \ref{def_tln}, becomes
\begin{align}
\tau_{0}\omega'\nu^{(1)}_{\textrm{Kerr(slow)}\,lm}&=\frac{\left(2+l\right)! \left(l-2\right)!\left(l!\right)^{2}}{\left(2 l\right)!\left(2l+1\right)!}\left[\frac{6mM\Omega_h}{2l+1}+2M\omega'\left\{1+\frac{3}{2l+1}\right\}\right]~,
\\
\tau_{0}\omega'\nu^{(2)}_{\textrm{Kerr(slow)}\,lm}&=\frac{\left(2+l\right)! \left(l-2\right)!\left(l!\right)^{2}}{\left(2 l\right)!\left(2l+1\right)!}\left[2M\omega'\left\{1+\frac{3}{2l+1}\right\}\right]~.
\end{align}
As is evident, like with the tidal Love numbers, the dissipative part also differs in the two approaches. The difference is solely due to keeping all the terms linear in $M\omega$, unlike the case of~\cite{Chia:2020yla}.  

\begin{table}[ht]
    \centering
    \setlength{\tabcolsep}{20pt}
    \renewcommand{\arraystretch}{1.5}
    \begin{tabular}{lll}
    \hline
        $(a/M)$ &  Response function of \cite{Chia:2020yla} & Response function from \ref{slow_fkerr} \\ \hline
        0 &  $(0 + 0.0666667 \,i) \,M\omega'$ & $(0 + 0.106667\, i) \,M\omega'$ \\
        
        0.01 &  $(0 + 0.0666667 \,i) \,M\omega'$ & $(0 + 0.0002\, i) + (0 + 0.106667\, i) \,M\omega'$ \\ 
        
       0.001 &  $(0 + 0.0666667 \,i) \,M\omega'$ & $(0 + 0.00002\, i) + (0 + 0.106667\, i) \,M\omega'$ \\  \hline
    \end{tabular}
\caption{Numerical estimations of the tidal response function associated with the gravitational wave mode ($l=m=2$) have been presented for a slowly rotating Kerr black hole. As is evident the response function derived in \cite{Chia:2020yla} differs from the expression derived here and most importantly has a nonzero part independent of the co-rotating frequency $\omega'$.}
\label{t2}
\end{table}

To summarize, the above result clearly shows that for nonaxisymmetric tidal perturbation---i.e., for modes with $m\ne0$---the zeroth-order term in the $M\omega'$ expansion of the response function of a slowly rotating black hole is nonzero. This is in sharp contrast with the corresponding result derived in \cite{Chia:2020yla}, where the response function was found to be proportional to $M\omega'$, and there was no zeroth-order term. This feature can also be seen from \ref{t2}, where we have provided numerical values of the response function for the $l=2=m$ mode with different choices of the dimensionless rotation parameter $(a/M)$. As is evident from \ref{t2}, the response function derived here, which includes all the terms $\mathcal{O}(M\omega)$, predicts a nonzero value for the zeroth-order term in the tidal response, which scales linearly with $(a/M)$. This is consistent with our findings---in particular, the result presented in \ref{slow_fkerr}. Moreover, the coefficient of $M\omega'$ differs from the expression of the tidal response function in \cite{Chia:2020yla}, which is clear from \ref{t2} and can be thought of as a generalization of the corresponding situation in the Schwarzschild spacetime, as seen in the previous section. If we interpret the zeroth-order term in the tidal response function as the Love numbers, then for a slowly rotating Kerr black hole, these are nonzero and imaginary.

\section{Discussion: The Love numbers of a rotating black hole}\label{discussion}

We have provided a detailed analysis involving the tidal response function of black holes under an external tidal field. Unlike the Newtonian analysis, we follow a covariant approach and define the tidal response function from the asymptotic expansion of the Weyl scalar $\psi_{4}$. Given the tidal response function, as elaborated earlier, there are two distinct ways of defining the tidal Love numbers from the tidal response function in the small-frequency approximation ($M\omega\ll1$). Among them, the first definition relates the tidal Love numbers with the real part of the response function, which can be straightforwardly applied to the case of an arbitrarily rotating black hole. The only nontrivial part is the determination of the response function, since the Teukolsky equation becomes complicated when all terms of $\mathcal{O}(M\omega)$ are included in the analysis (see \ref{App_C} for the solution of the Teukolsky equation in the context of an arbitrarily rotating black hole). Then the Love numbers simply follow by computing the real part of the response function.

However, the alternative way of defining the Love numbers, as advocated in \ref{def_tln}, may not even work for an arbitrarily rotating black hole. This is because the relation between the multipole moment and the tidal field cannot be truncated at the first-order time derivative in the corotating frame. In other words, in the Fourier space, the response function must depend on terms involving arbitrary powers of $M\omega'$. Then, expressing $\omega'$ in terms of $\omega$ and keeping terms up to linear order in $\omega$ should provide the Love numbers depending on arbitrary powers of the rotation parameter $a$. This clearly demonstrates that there are significant issues in obtaining the response function and then determining the Love numbers for an arbitrary rotating black hole if we follow the alternative approach, as discussed in \ref{Love_N_2}. Moreover, the series in \ref{def_I} may not even converge. Thus, the procedure of keeping all the powers of $\omega'$ and hence all the powers of the dimensionless rotation parameter $(a/M)$ in the tidal response function is not a feasible possibility. In particular, considering \ref{def_resp_sr} as an expansion in $\omega'$ is the crux of the trouble, as our analysis, based on the Teukolsky equation, works with the $M\omega\ll1$ approximation, and not for $M\omega'\ll 1$, and these two approximations do not coincide until and unless we content ourselves with the slow-rotation case. Therefore, for an arbitrarily rotating black hole, a small $M\omega$ does not imply $M\omega^\prime$ to be small, and hence we need to provide a way to circumvent this problem. Since we are working under the approximation $M\omega\ll1$, the most natural thing would be to expand the tidal response function in powers of the dimensionless quantity $M\omega$, which yields
\begin{equation}
F = F_0 + iF_1\,M\omega + \mathcal{O}(M^{2}\omega^{2})~.
\end{equation}
We can now replace $\omega$ with $\omega'+m\Omega_{\rm h}$, and hence the above response function can be expressed as a leading-order term and then a term linear in $\omega'$, such that
\begin{equation}
F = (F_0 + imM\Omega_h\,F_1) +iF_1\,M\omega'+\mathcal{O}(M^{2}\omega^{2})~.
\end{equation}
Having determined the term independent of $\omega'$ and the term linearly dependent on $\omega'$, we can propose the zeroth-order term to be related to the Love numbers, as in \ref{Love_N_2}, while the coefficient of the term $M\omega'$ is related to tidal dissipation, given by \ref{dissip_2}. As is evident, with this definition, the tidal Love numbers are manifestly imaginary, as long as the tidally deformed object is rotating, and hence does not represent a conservative quantity. When expressed explicitly, the tidal Love numbers read
\begin{equation}
k^{\rm (2)}_{lm} = \frac{1}{2}(F_0 +imM\Omega_h\,F_1)~,
\end{equation}
and the dissipative part becomes
\begin{equation}
\tau_0\nu^{\rm (2)}_{lm} = F_1\,M~.
\end{equation}
Note that unlike \ref{def_resp_sr}, the above expression for the response function is a series in the frequencies observed by an asymptotic observer, rather than the frequencies measured in the corotating frame of reference. It is clear that finding the part of the response function which is independent of $\omega'$ is a challenging task, and in general can lead to imaginary tidal Love numbers, completely counterintuitive to its conservative nature. Thus, defining the tidal Love numbers from the real part of the response function seems to be the way forward, since it is not ambiguous in the presence of an arbitrarily rotating black hole, nor it is imaginary at any level. 

Despite the above difficulties and subtleties in defining the Love numbers of an arbitrarily rotating black hole, we believe that some significant results have been obtained in this work. First of all, we have provided a relativistic definition of the tidal response function from the Weyl scalar $\psi_4$ and subsequently have explicitly described the ambiguities and inconsistencies in defining the Love numbers from the tidal response function, for the first time. In particular, we have put forward two possible ways of addressing them. Second, we have explicitly pointed out how the small-frequency approximation should be taken, and what corresponds to the structure of the radial Teukolsky equation. It turns out that earlier results in this direction had missed quite a few terms linear in $M\omega$, and thus our results derived here are complete, in the sense that they encapsulate effects from all terms of $\mathcal{O}(M\omega)$. Following the modified radial Teukolsky equation at our disposal, we have calculated the response functions of nonrotating as well as slowly rotating black holes in the near-horizon and small-frequency limit, while keeping track of all the terms linear in $M\omega$. As emphasized earlier, this is unlike the earlier works such as Ref.~\cite{Chia:2020yla}. Our results suggest vanishing tidal Love numbers for a Schwarzschild black hole; however, they also predict that the dissipative effects for the Schwarzschild black hole will be stronger than what has been obtained earlier~\cite{Chia:2020yla}. In contrast, the Love numbers for a slowly rotating black hole in a tidal environment depend on the details of the procedure used to define them, as was described in this paper: While one procedure yields vanishing Love numbers, the other suggests nonzero, but purely imaginary, Love numbers. Our results regarding the tidal Love numbers as well as the dissipative terms provide corrections over and above the recent results derived in \cite{Chia:2020yla}, due to the inclusion of all the linear-order terms of $\mathcal{O}(M\omega)$, some of which were missing in \cite{Chia:2020yla}. 

This work can also be extended by including higher-order terms in $M\omega$---to be precise, second order or more---in the Teukolsky equation, thereby calculating the response function. It would also be interesting to see how the above formalism can be used in determining the tidal response function of exotic compact objects and possibly those of quantum black holes, so that we have an understanding of the tidal Love numbers for these objects. We also hope to arrive at the expression for the tidal Love numbers and the tidal dissipation for an arbitrary rotating black hole, from our proposed procedure presented here, elsewhere. Finally, it will be helpful to entangle the response function of black holes, or compact objects in general, to the inspiral part of the gravitational wave signal through the Weyl scalar. This will allow us to directly relate the tidal Love numbers and the tidal dissipation to the gravitational wave waveform, which in turn can possibly make the smoking-gun test regarding the nature of the coalescing compact objects more feasible. 

\section*{Acknowledgements}

The research of S. C. is funded by the INSPIRE Faculty fellowship from the DST, Government of India (Reg. No. DST/INSPIRE/04/2018/000893). A part of this work was performed at the Albert Einstein Institute under the Max-Planck-India Mobility Grant from the Max-Planck Society, and S. C. acknowledges the warm hospitality at the Albert Einstein Institute. We also acknowledge support from the NSF under Grant No. PHY-2309352. This document has been assigned the LIGO document number LIGO-P2300180.

\appendix
\labelformat{section}{Appendix #1} 
\labelformat{subsection}{Appendix #1}
\section{Master equation from the radial Teukolsky equation}\label{app_A}

In this section, we shall show the calculations involved to arrive at \ref{app_gen_eq}, which serves as the master equation for this work, from the Teukolsky equation presented in \ref{chiaA41}. Under the transformation $z=(r-r_+)/(r_+-r_-)$, the Teukolsky equation in \ref{chiaA41} can be written as
\begin{multline}
    \frac{\mathrm{d}^2R}{\mathrm{d}z^2} + \left\{\frac{2iP_+-1}{z}-\frac{2iP_-+1}{z+1}-2i\omega(r_+-r_-)\right\}\frac{\mathrm{d}R}{\mathrm{d}z} \\ +\left\{\frac{4iP_-}{(z+1)^2}-\frac{4iP_+}{z^2}+\frac{A_-+iB_-}{(1+z)}-\frac{A_++iB_+}{z}\right\}R=\frac{T}{\Delta}(r_+-r_-)^2~,
\end{multline}
which can also be expressed as
\begin{multline}
    \frac{\mathrm{d}^2R}{\mathrm{d}z^2} + \left\{\frac{2iP_+-1}{z}-\frac{2iP_-+1}{z+1}-2i\omega(r_+-r_-)\right\}\frac{\mathrm{d}R}{\mathrm{d}z} + \left\{\frac{4iP_-}{(z+1)^2}-\frac{4iP_+}{z^2}-\frac{l(l+1) -2}{z(1+z)} \right.\\+\left.\frac{2ma\omega}{z(1+z)}\left\{1+\frac{4}{l(l+1)}\right\} -\frac{2i\omega r_+}{z(z+1)}- \frac{2i\omega}{z+1}(r_+-r_-) +\mathcal{O}[(a\omega)^2]\right\}R=\frac{T}{\Delta}(r_+-r_-)^2~.
\label{transformed_eq_2}
\end{multline}
We will now simplify the coefficients of $(\mathrm{d}R/\mathrm{d}z)$ and $R$.

\subsection{Coefficient of \texorpdfstring{$(\mathrm{d}R/\mathrm{d}z)$}{TEXT}}

Using the near-horizon and the small-frequency approximation $(M\omega\ll1)$, we can neglect $2i\omega(r_+-r_-)z/(z+1)$ in the coefficient of $\frac{\mathrm{d}R}{\mathrm{d}z}$, which implies
\begin{equation}
    \frac{2iP_+-1}{z}-\frac{2iP_-+1}{z+1}-2i\omega(r_+-r_-) \sim \frac{2iP_+-1}{z} -\frac{2iP_1+1}{z+1}~,
\end{equation}
where $P_- + \omega (r_+-r_-) = P_1$.

\subsection{Coefficient of \texorpdfstring{$R$}{TEXT}}

Using the near-horizon and the small-frequency approximation $(M\omega\ll1)$, we can neglect $2i\omega(r_+-r_-)z/(z+1)^2$ and $\mathcal{O}[(a\omega)^2)]$ terms in the coefficient of $R$, yielding
\begin{equation}
\frac{4iP_-}{(z+1)^2}-\frac{2i\omega}{z+1}(r_+-r_-) + \mathcal{O}[(a\omega)^2)]\sim \frac{4iP_2}{(z+1)^2}~,
\end{equation}
where $P_- -\frac{1}{2} \omega (r_+-r_-) = P_2$. Now, we can obtain \ref{app_gen_eq} from \ref{transformed_eq_2}, by substituting $T=0$ as well as the simplified coefficients of $(dR/dz)$ and $R$ in the relevant equation.

\section{Calculation of the tidal response function}\label{app_B}

In this appendix, we depict the computation of the tidal response function by expanding out the gamma functions appearing in \ref{resp_Sch} for the Schwarzschild black hole, and \ref{resp_Kerr} for the slowly rotating Kerr black hole. In general, the response function takes the following form: 
\begin{equation}
F = \frac{\Gamma\left(1 + l + 2 i P_+ + Q_1\right) \Gamma\left(3 + l - Q_2\right) \Gamma\left(-1 - 2 l - Q_1 + Q_2\right)}{ \Gamma\left(1 + 2 l + Q_1 - Q_2\right)\Gamma\left(2 - l - Q_1\right) \Gamma\left(-l + 2 i P_+ + Q_2\right)}\,,
\end{equation}
where $P_+$, and $Q_{1,2}$ are all small quantities in our approximations for both the Schwarzschild and the slowly rotating Kerr black holes. Since our analysis is only valid up to the first order in $M\omega$ (and $a/M$ for the slowly rotating Kerr black hole); therefore, we will neglect all second- and higher-order terms in the gamma functions appearing in the tidal response function, presented above.

For this purpose, let us consider a function $F(z)$; we can expand it around $z=z_0$ using the following Taylor expansion
\begin{equation}
F(z)=F(z_0) + (z-z_0)\frac{\mathrm{d}F}{\mathrm{d}z}\bigg|_{z=z_0} + \mathcal{O}[(z-z_0)^2]~.
\end{equation}
Similar expansion can be obtained for a gamma function---say, $\Gamma(f(z))$---which yields
\begin{multline}
\Gamma(f(z))=\Gamma(f(z_0)) + (z-z_0)\frac{\mathrm{d}\Gamma(f(z))}{\mathrm{d}z}\bigg|_{z=z_0} + \mathcal{O}[(z-z_0)^2] \\=\Gamma(f(z_0)) + (z-z_0)\psi(f(z_0))\Gamma(f(z_0))\frac{\mathrm{d}f(z)}{\mathrm{d}z}\bigg|_{z=z_0} + \mathcal{O}[(z-z_0)^2]~,
\end{multline}
where $\psi(z)$ is the digamma function, defined as \cite{abramowitz_stegun_1972}
\begin{equation}
\psi(z) = \frac{1}{\Gamma(z)}\frac{\mathrm{d}\Gamma(z)}{\mathrm{d}z}~.
\end{equation}
Expanding out each and every gamma functions in the definition of the response function, and neglecting the higher-order terms, we obtain
\begin{equation}
F = \frac{\Gamma\left(1 + l\right) \Gamma\left(3 + l\right) \Gamma\left(-1 - 2 l\right)}{ \Gamma\left(1 + 2 l\right)\Gamma\left(2 - l\right) \Gamma\left(-l\right) }\left[1+A_1+A_2\right]~,
\end{equation}
where 
\begin{multline}
A_1= (2 i P_+ + Q_1)\psi(1 + l)+(- Q_2)\psi(3 + l)+(- Q_1 + Q_2)\psi(2+2l)\\ -(Q_1 - Q_2)\psi(1 + 2 l)+ Q_1\psi(-1+l) - (2 i P_+ + Q_2)\psi(1+l)~,
\end{multline}
and
\begin{equation}
A_2= (- Q_1 + Q_2)\pi\cot(2l\pi)- (2 i P_+ + Q_2-Q_1)\pi\cot(l\pi)~.
\end{equation}
In arriving at the above result, we have used the identity $\psi(1-z) = \psi(z) +\pi\cot(\pi z)$~\cite{abramowitz_stegun_1972}. For $l\in\mathbb{L}=\mathbb{Z}^+\backslash\{1\}$, we obtain
\begin{equation}
\lim_{l\rightarrow l}\frac{\Gamma\left(1 + l\right) \Gamma\left(3 + l\right) \Gamma\left(-1 - 2 l\right)}{ \Gamma\left(1 + 2 l\right)\Gamma\left(2 - l\right) \Gamma\left(-l\right) } = 0~,
\end{equation}
and
\begin{equation}
\lim_{l\rightarrow l}\frac{\Gamma\left(1 + l\right) \Gamma\left(3 + l\right) \Gamma\left(-1 - 2 l\right)}{ \Gamma\left(1 + 2 l\right)\Gamma\left(2 - l\right) \Gamma\left(-l\right) }A_1 = 0~,
\end{equation}
along with
\begin{equation}
\lim_{l\rightarrow l}\frac{\Gamma\left(1 + l\right) \Gamma\left(3 + l\right) \Gamma\left(-1 - 2 l\right)}{ \Gamma\left(1 + 2 l\right)\Gamma\left(2 - l\right) \Gamma\left(-l\right) }A_2 =-\frac{\Gamma\left(1 + l\right) \Gamma\left(3 + l\right) \Gamma\left(l-1\right)\Gamma\left(1+l\right)}{2\Gamma\left(1 + 2 l\right)\Gamma\left(2l+2\right)}\left[2iP_+ +\frac{1}{2}(Q_2-Q_1)\right]~.
\end{equation}
Collecting all these results, for $l\in\mathbb{L}=\mathbb{Z}^+\backslash\{1\}$, we obtain the following result for the tidal response function:
\begin{equation}
F = -\frac{1}{2}\frac{\Gamma\left(1 + l\right) \Gamma\left(3 + l\right) \Gamma\left(l-1\right)\Gamma\left(1+l\right)}{\Gamma\left(1 + 2 l\right)\Gamma\left(2l+2\right)}\left[2iP_+ +\frac{1}{2}(Q_2-Q_1)\right]~,
\end{equation}
which reduces to \ref{resp_Sch_1} and \ref{resp_Kerr_1} for the Schwarzschild and the slowly rotating Kerr black hole, respectively.

\section{General solution of the Teukolsky equation in the small-frequency limit}\label{App_C}

We have explicitly provided the solution of the Teukolsky equation in the small-frequency limit, for nonrotating as well as for slowly rotating black holes. In this appendix we provide the general solution of the Teukolsky equation in the small-frequency limit for arbitrarily rotating black holes. The corresponding equation in the small-frequency limit has already been presented in \ref{app_gen_eq}, whose solution for an arbitrarily rotating black hole reads
\begin{multline}
    R(z) =  z^2 (1 + 
  z)^{H_1}  C_2 \,F\left(\frac{1}{2} (3 + 2 i P_+ + i D_1), I_1; 
     3 + 2 i P_+; -z\right)
     \\
  +z^{-2 i P_+} (1 + 
  z)^{H_1} C_1 \,F\left(\frac{1}{2}  (-1 - 2 i P_+ + i D_1), 
     I_1 - 2 -2 i P_+; -1 - 2 i P_+; -z\right)~,
\end{multline}
where $C_1$ and $C_2$ are constants arising from the second-order Teukolsky equation. Since our analysis is valid only up to linear orders of $M\omega$, here also we have expanded the quantities $D_1$, $I_1$, and $H_1$ in the linear orders of $M\omega$ (neglecting second- and higher-order term), yielding the following expression for the quantity $D_{1}$:
\begin{multline}
    D_1 = \frac{l (1 + l) (-i + 4 i l (l + 5 i M\omega) + 14 M\omega) + 
  2 i (-8 + l (1 + l) (1 + 6 l)) m a\omega}{l (1 + l) (1 + 2 l)} 
  \\ 
  +\frac{2 M^2\omega - m a}{r_+-M} + \frac{16 (-1 + l) r_+\omega}{1 + 2 l} - \frac{
 6 m^2 a^2 \omega}{2 r_+ - i m a -2M }\,.
\end{multline}
Along identical lines, the expression for $I_{1}$ reads
\begin{multline}
    I_1 = 3+l +\frac{l (1 + l) ( - 6 i (3 + 2 l) M\omega) - (8 + 
     l (1 + l) (5 + 6 l)) m a\omega}{l (1 + l) (1 + 2 l)} 
     \\
     +\frac{
 3 m^2 a^2\omega }{2ir_+ -2iM + m a} + \frac{8 i (2 + l) r_+\omega}{
 1 + 2 l}\,,
\end{multline}
and finally, for the remaining quantity $H_{1}$, we obtain
\begin{equation}
    H_1 = 2-\frac{ 12 (r_+-M)^2\omega}{2ir_+ -2iM +ma}~.
\end{equation}
These expressions will be used in the future for determining the response function of an arbitrarily rotating black hole, from which an expression for the tidal Love numbers can also be arrived at.
\bibliographystyle{apsrev4-1}
\bibliography{reference}
\end{document}